\documentclass[12pt]{article}
\usepackage[utf8]{inputenc}

\usepackage[square,numbers, sort&compress]{natbib}
\usepackage{times}

\usepackage[margin=1in]{geometry}
\usepackage{graphicx}
\usepackage{color}

\usepackage{paralist}

\newenvironment{innerlist}[1][\enskip\textbullet]%
        {\begin{compactenum}[#1]}{\end{compactenum}}

\begin{document}


\vspace*{2cm}

\thispagestyle{empty}

\begin{center}

{\Large Multipoint Measurements of the Solar Wind: \\A Proposed Advance for
Studying Magnetized Turbulence}

\vspace*{1cm}

{\large
K.G. Klein\\
University of Arizona\\
Tel: 520-621-2806, Email: kgklein@email.arizona.edu\\

\vspace*{1cm}

Co-Authors:\\
O. Alexandrova$^{1}$,
J. Bookbinder$^{2}$,
D. Caprioli$^{3}$,
A.W. Case$^{4}$, 
B.D.G. Chandran$^{5}$,\\
L.J. Chen$^{6}$,
T. Horbury$^{7}$, 
L. Jian$^{6}$, 
J.C. Kasper$^{8}$, 
O. Le Contel$^{9}$, 
B.A. Maruca $^{10}$,\\
W. Matthaeus $^{10}$, 
A. Retino$^{9}$,
O. Roberts $^{11}$, 
A. Schekochihin$^{12}$, 
R. Skoug$^{13}$,\\
C. Smith$^{5}$,
J. Steinberg$^{13}$,
H. Spence$^{5}$, 
B. Vasquez$^{5}$,
J. M. TenBarge $^{14}$,\\
D. Verscharen$^{5,15}$,
P. Whittlesey$^{16}$

\vspace*{1cm}
$^{1}$Observatoire de Paris, LESIA
$^{2}$NASA Ames
$^{3}$University of Chicago
$^{4}$Smithsonian Astrophysical Observatory
$^{5}$University of New Hampshire
$^{6}$NASA/GSFC
$^{7}$Imperial College
$^{8}$University of Michigan
$^{9}$Laboratorie de Physique des Plasmas
$^{10}$University of Delaware
$^{11}$Space Research Institute, Austrian Academy of Sciences
$^{12}$University of Oxford 
$^{13}$LANL\\
$^{14}$Princeton University
$^{15}$University College London\\
$^{16}$U.C. Berkeley Space Sciences Lab
}

\end{center}

\newpage

\setcounter{page}{1}

A multi-institutional, multi-national science team will soon submit a NASA proposal to build a swarm of spacecraft to fly into the near-Earth solar wind in a configuration spanning a multitude of scales in order to obtain critically needed measurements that will reveal the underlying dynamics of magnetized turbulence. This white paper provides a brief overview of turbulent systems that constitute an area of compelling plasma physics research, including why this mission is needed, and how this mission will achieve the goal of \textbf{revealing how energy is transferred across scales and boundaries in plasmas throughout the universe.}\\

\textbf{The Motivation for Studying Magnetized Turbulence}
Plasmas are a ubiquitous state of matter. Turbulence, the nonlinear scale-to-scale transfer of energy, plays a critical role in the transport of mass, momentum, and energy in plasma systems as varied as solar and stellar winds, black hole accretion disks, the interstellar media, and terrestrial laboratory environments.
The motivation to understand turbulence is both diverse and fundamental.
The general characterization of turbulence remains one of the outstanding tasks of classical physics, in particular understanding how energy is transported through a system, and how such transport changes the evolution of the system. 
Turbulence is fundamental to understanding solar wind plasma acceleration and heating, the nature of plasma fluctuations everywhere, and the scattering, acceleration, and transport of energetic particles. 
Progress on characterizing turbulence in plasma systems will improve our understanding of some of the most important processes in astrophysics, including the formation of stars and planets, the heating of accretion flows, and turbulent dynamo generation. 
While significant theoretical effort has been expended, there are still open questions about the mechanisms that transfer energy from scale-to-scale within the turbulent cascade as well as those that dissipate cascade energy and energize the particles. 
There is no universal agreement on the spectral or spatial distribution of turbulent power, with different theories predicting different distributions of power. 
Characterizing these properties will enable the description of the thermodynamic fate of plasmas throughout the universe.

\textbf{What is Unknown about Plasma Turbulence?}
While there has been significant progress in understanding turbulence over the last eight decades, there are still a number of vital, unanswered questions about the nature of plasma turbulence.
In particular, the nature of the turbulent fluctuations, the rate at which energy is transferred, the role of magnetic fields in organizing the transfer, how the plasma changes as it transitions from magnetofluid to kinetic scales, and the mechanisms that remove energy from the turbulent cascade are all topics of intense research.
A review of some of the theories proposed to address these questions can be found in white papers submitted to the 2020 Plasma Decadal Panel by TenBarge et al \cite{TenBarge:2019}and Matthaeus et al \cite{Matthaeus:2019}.

One of the key challenges to modeling turbulence is the vast spatial and temporal scales covered by most turbulent systems.  
The length scales at which energy is injected into a turbulent system and those at which they are removed are typically separated by several orders of magnitude, with physical processes having characteristic timescales spanning similar separations.
Any numerical simulation of such systems must choose between realistic scale separations, which are necessary to generate appropriate nonlinear structures, and kinetic plasma mechanisms, which are necessary to accurately damp and dissipate the energy from the cascade onto the ions and electrons. 
Similarly, laboratory experiments are of an insufficient size to generate a large enough inertial range to replicate solar and astrophysical cascades, and astrophysical systems are too remote to perform in situ diagnostics of turbulent phenomena.

The solar wind offers the most accessible environment for the in situ observation at all relevant scales of turbulent electromagnetic fields and particle distributions that are representative of magnetized turbulence throughout the universe\cite{Bruno:2013}.
At the largest scales, greater than $10^6$ km, solar wind varies as a result of velocity shear, compressions, and magnetic boundaries that are a direct result of solar structure and activity. 
This constitutes the \textit{energy-containing range} that drives the turbulence at smaller scales. 
Between $10^3$ and $10^6$ km, the \textit{inertial range} contains the nonlinear dynamics that transport energy from larger to smaller scales. 
This part of the spectrum often exhibits a universal form, typically a power-law spectrum, that is indicative of the nonlinear dynamics. 
At scales smaller than $10^3$ km, plasma kinetics scales are reached, leading to dissipation as the energy in the turbulent fluctuations is converted into thermal energy of the ions and electrons.

To date, all in situ observations of solar wind plasmas have been single point measurements (e.g., ACE or Wind)\cite{Harten:1995,McComas:1998a}, or have focused on a narrow range of scales through the use of carefully controlled formations of a few spacecraft (e.g., Cluster, MMS, or THEMIS)\cite{Escoubet:2001,Burch:2016,Angelopoulos:2008}.
These missions have been successful, with Wind alone having 4615 peer-reviewed publications from 1995 to 2017\cite{Wind}, but they are fundamentally limited in their application to studying turbulence.
Single-point observations must map time series to spatial structure through Taylor's hypothesis\cite{Taylor:1938}, which assumes advected time scales are much faster than any temporal evolution in the plasma frame. 
This hypothesis is frequently invoked, but may not be valid at the same scales where energy is removed from the turbulent cascade.
Even under conditions where Taylor's hypothesis is valid, one can only sample spatial structure parallel to the flow direction. 
A central problem in understanding turbulence is that key dynamics, both linear and nonlinear, depend upon the orientation of the wave vector relative to the mean magnetic field; no single spacecraft can measure this three-dimensional quantity.
Assembling information on the three-dimensional structure of turbulence with a single point measurement requires averaging over long intervals with different flow directions, potentially mixing different kinds of turbulence together.

These limitations of single-point measurements are widely recognized and served as the impetus for the CLUSTER and MMS magnetospheric missions.
Flying four spacecraft in a tetrahedral formation, as was done with these missions, provides a limited set of separations between the spacecraft that can be used to sample the multi-directional structure of the plasma. 
In the case of MMS, these separations allow the study of the multi-dimensional dynamics of magnetic reconnection at small scales.
Highlights of the turbulence science that have been extracted from such single-scale missions are discussed in a white paper submitted to the 2020 Plasma Decadal Panel by Chen et al\cite{Chen:2019}.
At a given time, the inter-spacecraft separations typically have very similar lengths, and therefore are inadequate to simultaneously resolve the cross-scale nonlinear couplings needed to understand the dynamics of plasma turbulence.
Even with the most advanced analysis techniques, the scales sampled only cover a factor of approximately ten \cite{Sahraoui:2010}, nowhere near the orders of magnitude necessary for simultaneously measuring turbulent fluctuations through the inertial and dissipation ranges. 

As turbulence is fundamentally a multi-scale, three-dimensional, time-evolving phenomenon, neither single-point measurements nor even a cluster of four spacecraft can provide the measurements necessary to reach closure on fundamental outstanding questions in plasma turbulence.
To reveal the entire temporal and spatial structure of turbulence requires observations at an array of points that far exceed the single tetrahedral configurations flown to date.
\textbf{Making progress on fundamental questions about the multi-scale and three-dimensional nature of turbulence requires simultaneous measurements from a large number of spacecraft at a variety of separation distances.}

\textbf{HelioSwarm: An Advance for Studying Magnetized Plasma Turbulence}
The goal of the proposed mission concept, named HelioSwarm\cite{HelioSwarm}, is to {reveal how energy is transferred across scales and boundaries in turbulent plasmas throughout the universe.}
This task will be accomplished by flying a swarm of small spacecraft with a wide range of spatial separations to simultaneously sample the key physical parameters in the turbulent solar wind.
The newly-feasible, cost-effective mission will reveal and quantify key, currently unmeasured aspects of turbulence, allowing us to describe the cascade of energy across scales and into different physical regions, where energy is dissipated.
This mission will also provide a means of directly testing current conflicting models for the spectral and spatial distributions of turbulent power, which in turn affects our understanding of dissipation and scattering.
Specifically, this mission will 
\begin{innerlist}
    \item Reveal the three-dimensional spatial and temporal distribution of turbulence in plasmas.
    \item Unveil the thermodynamic pathways to dissipation.
    \item Ascertain the mutual impact between boundaries and large-scale structure and turbulence.
\end{innerlist}


Comparison with previous multispacecraft missions, such as MMS,
suggests that HelioSwarm will be prohibitively expensive; this is not
the case.  Where MMS, as well as previous multiscale proposals such as
CrossScale\cite{CrossScale}, seek to tightly control the orbits of
four or more extensively instrumented spacecraft in tight formation,
the HelioSwarm Observatory will fly eight more simply instrumented
spacecraft in a loose formation around a central hub with constantly
changing inter-spacecraft separation, thereby forming and reforming
desirable configurations on multiple scales.  The eight-node
spacecraft and the hub will measure the vector magnetic field with
fluxgate and searchcoil magnetometers, and proton properties with
Faraday cups.  The hub spacecraft will have additional electrostatic
analyzers to measure velocity distribution functions in more detail
and will be responsible for collecting and downlinking data from the
swarm.  All of the proposed instruments have heritage on current and
upcoming NASA and ESA missions, decreasing cost.

\begin{figure}[b]
    \centering
{\includegraphics[width=1.0\textwidth]{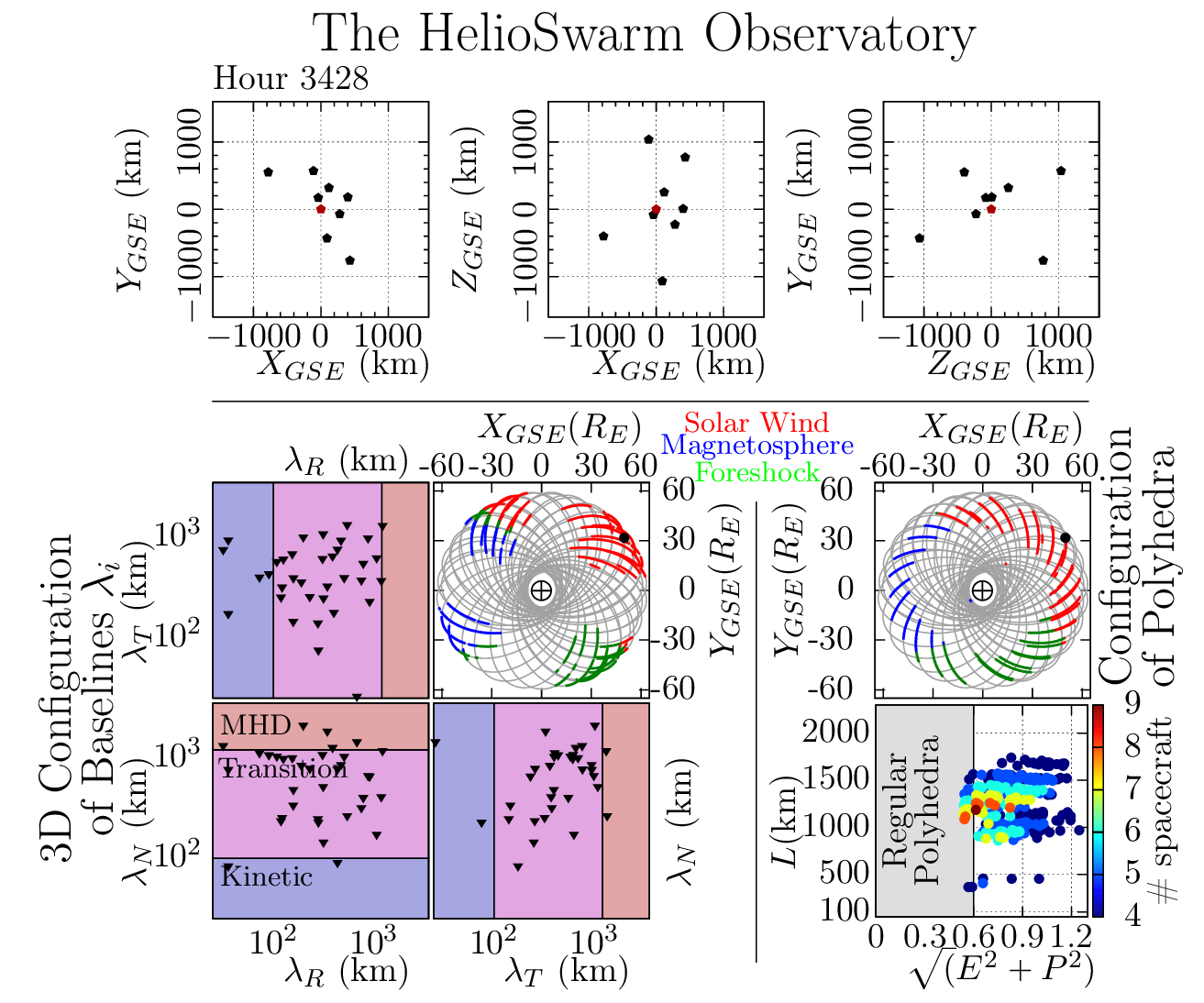}}
\caption{Summary description of an example HelioSwarm orbit.
  \textbf{Top:} Relative distances between nodes (black) and hub (red)
  in GSE co\"ordinates at 3428 hours into a sample orbit.
  \textbf{Lower Left:} The RTN components of the 36 baselines
  $\mathbf{\lambda}$ between the nine spacecraft, demonstrating good
  three-dimensional coverage of MHD (red), transition (purple), and
  kinetic (blue) scales at the selected hour.  The near-Earth regions
  where such coverage is satisfied are shown in color (red: pristine
  solar wind; blue: magnetosphere; green: foreshock) with the
  observatory orbit shown in grey and current location in black.
  \textbf{Lower Right:} Average inter-spacecraft distances $L$
  vs. shape (sum of elongation $E$ and planarity $P$) of the 382
  polyhedra composed of at least four HS spacecraft; color indicates
  the number of vertices.  Near-Earth regions with at least two
  polyhedra with $\sqrt{E^2+P^2}\leq 0.6$ that satisfy $L_1 \geq 3
  L_2$ are highlighted in color in the same format as the $\lambda$
  plot.  A movie illustrating the the evolution of the observatory is
  included in the ancillary materials and at Ref.~\cite{HelioSwarm}. }
    \label{fig:example}
\end{figure}

A summary plot of an example orbit for the HelioSwarm Observatory is
shown in Fig.~\ref{fig:example}, as well as in the ancillary movie
file, which can also be found at Ref.~\cite{HelioSwarm}. The
Observatory will travel in a lunar-resonant orbit optimized to
minimize station-keeping, sampling pristine solar wind, as well as the
magnetosphere, magnetosheath, and magnetically connected solar wind.
A typical distribution of the nodes with respect to the hub at hour
3428 is shown in the top row of Fig.~\ref{fig:example}; the position
of HS with respect to the Earth is indicated by a black circle middle
row panels.

The thirty-six baseline distances $\mathbf{\lambda}$ separating the
eight nodes and single hub will simultaneously span MHD, transition,
and ion kinetic scales.  As ion kinetic scales, e.g. proton inertial
lengths and gyroradii, are on the order of 100 km in the pristine
solar wind, minimum baselines will be as small as 50 km.  Since the
break scale between MHD scales and the dissipation range is typically
found to be on the order of 1000km, it will be necessary to
simultaneously measure baselines as large as 3000 km.  Because of the
natural anisotropies in the system due to the solar wind flow
direction and the local magnetic field, it is essential that the
baselines not be aligned in a single line or plane, but rather cover
baselines along and across the local flows and fields so that the
three-dimensional structure of solar wind turbulence can be
reconstructed. Such good three-dimensional coverage is demonstrated in
the lower-left corner of Fig.~\ref{fig:example}; the projections of
$\mathbf{\lambda}$ onto the RTN axis simultaneously cover MHD,
transition, and kinetic scale ranges in all three orthogonal
directions.  Measurements at these spatially separated locations will
enable the first ever multi-scale calculation of two-point
correlations with time and space as independent inputs, as well as
multi-scale studies of cascade rates and intermittancy.

Using at least four of the nine spacecraft from the observatory, 382
distinct polyhedra can be constructed. These polyhedra will allow for
the study of the three-dimensional structure of the solar wind
turbulence.  HelioSwarm will have an enormous range of average
interspacecraft spacings, as well as varying associated elongations
and planarities, plotted in the lower-right corner of
Fig.~\ref{fig:example}, allowing it to take advantage of the numerous
multi-spacecraft analysis techniques developed for formation flying
\cite{Paschmann:2008}, including direct calculation of spatial
derivatives as well as those capable of measuring the distribution of
power in frequency-wavevector space as a function of size scale and
orientation.  Combined with multi-point correlation studies, these
techniques enable detailed studies of the distribution, transport, and
removal of energy in solar wind turbulence in a way inaccessible to
any current or previous mission.

With the advent of low-resource sensors and small satellites, configurations of many spacecraft simultaneously sampling many scales are possible for the first time and promise to transform our knowledge of turbulence. \textbf{A mission like the soon-to-be proposed HelioSwarm concept is necessary to bring closure to some of the most pressing open questions in the study of solar wind, and advance our understanding of plasma turbulence throughout the universe.}

\vspace*{-1.0cm}

\bibliographystyle{apsrev} 
\bibliography{main}

\end{document}